\documentclass[aps,prl,twocolumn,superscriptaddress,preprintnumbers]{revtex4-2} % PRL standard class

\usepackage{amsmath,amssymb,amsfonts} 
\usepackage{graphicx} 
\usepackage{hyperref}
\usepackage{bm} 
\usepackage{times} 
\usepackage{xcolor}
\usepackage{physics}
\usepackage{multirow}
\usepackage{pifont}
\usepackage{braket}

\usepackage{tikz}

\definecolor{darkgreen}{HTML}{006400}

\begin{document}

\title{Non-Local Magic from the Entanglement Spectrum}
	
\author{Gianpaolo Torre}
\affiliation{Jozef Stefan Institute, Jamova cesta 39, 1000 Ljubljana, Slovenia}
	
\author{Fabio Franchini}
\affiliation{Institut Ruder Bošković, Bijenička cesta 54, Zagreb 10000, Croatia}
	
\author{Salvatore Marco Giampaolo}
\affiliation{Institut Ruder Bošković, Bijenička cesta 54, Zagreb 10000, Croatia}

\begin{abstract}
Non-local magic has recently emerged as a fundamental resource for characterizing genuinely non-local, basis-independent, non-stabilizer correlations.
However, its evaluation is computationally intractable beyond small systems, and its analytical properties remain largely unexplored.
We show that the Schmidt-gauge formulation of non-local magic admits an equivalent representation in terms of the Walsh–Hadamard autocorrelations of the entanglement spectrum, thereby making its hidden harmonic structure explicit.
This formulation enables a systematic analytical treatment, allowing us to derive exact results for broad classes of quantum states, establish a universal upper bound in terms of the second-order Rényi entanglement entropy, obtain a perturbative estimation in typical volume-law states and in the ground states of one-dimensional gapped systems, and uncover a double-logarithmic scaling scenario at one-dimensional criticality, emerging from the universal Calabrese–Lefevre entanglement spectrum and supported by numerical calculations.
Our results establish entanglement spectroscopy as a natural analytical framework for investigating non-local magic, while identifying the Walsh autocorrelation structure of the entanglement spectrum as its fundamental organizing principle.
\end{abstract}

\preprint{RBI-ThPhys-2026-25}
	
\maketitle

Understanding the complexity of quantum many-body states is one of the central challenges of modern quantum physics. 
Over the past couple of decades, quantum entanglement has profoundly reshaped our understanding of this problem,
providing a common framework for phenomena as diverse as quantum criticality, topological phases of matter, tensor-network methods, and quantum dynamics~\cite{Vidal2003, Hamma2005, Amico2008, Eisert2010, Orus2014, Laflorencie2016}.
The entanglement entropy and, more generally, the entanglement spectrum have revealed universal structures that are largely independent of microscopic details, establishing entanglement as one of the fundamental organizing principles of modern many-body physics~\cite{Hamma2005_2,LiHaldane2008,Calabrese2008,Peschel2009}.

Nevertheless, entanglement alone captures only one aspect of quantum complexity.
While it quantifies the amount of bipartite quantum correlations shared between two subsystems, it does not distinguish between correlations that admit an efficient stabilizer description and those requiring genuinely non-stabilizer resources.
The latter are commonly referred to as {\it ``quantum magic''}~\cite{Gottesman1998}. 
This distinction has become increasingly important with the development of quantum information theory, where quantum magic has been revealed as the fundamental resource underlying universal fault-tolerant quantum computation, quantum advantage, and the classical hardness of quantum simulation~\cite{BravyiKitaev2005, Veitch2014, Howard2014, Leone2022}.
More recently, magic has also attracted considerable attention in quantum many-body physics, where it has been shown to reveal physical properties that remain invisible to conventional entanglement measures, establishing new connections between quantum information, condensed matter physics, and quantum complexity~\cite{Oliviero2022, Turkeshi2026, Odavic2023, Catalano2025}.

Among different notions of magic, the non-local magic introduced in Ref.~\cite{Cao2025} provides a particularly natural measure of genuinely basis-independent bipartite non-stabilizer resources. 
By construction, it removes all contributions that can be generated by local unitary transformations and isolates only the irreducible many-body component of the resource~\cite{Liu2026}. 
This makes it a natural quantity for characterizing the intrinsic complexity of bipartite quantum states.

Despite its conceptual importance, our analytical understanding of non-local magic remains remarkably limited. 
Its original definition involves a highly non-trivial and generally intractable optimization over local unitary transformations and, 
despite several attempts, the mathematical structure underlying the resulting spectral functional remains largely unexplored. 
Consequently, some of the most relevant questions remain essentially unanswered. 
What determines the scaling of non-local magic? 
What is its precise relation to entanglement? 
Which properties of the entanglement spectrum are responsible for large non-local magic? 
Does non-local magic exhibit universal behavior in quantum critical systems?

With this work, we answer these questions, by showing that the Schmidt-gauge formulation of non-local magic, already introduced in~\cite{Cao2025}, admits an exact representation in terms of the Walsh--Hadamard autocorrelations of the entanglement spectrum.
This formulation makes the underlying harmonic structure explicit and naturally lends itself to analytical investigations~\cite{NielsenChuang}.
It enables us to derive exact analytical results for several physically relevant families of quantum states, establish a universal upper bound in terms of the second-order Rényi entanglement entropy, and identify the spectral mechanisms responsible for qualitatively different behaviors of non-local magic.
In particular, we characterize non-local magic of maximally entangled states, Haar-random states, factorized entanglement spectra, and one-dimensional gapped systems, while revealing the role of Schmidt ordering in determining the scaling of non-local magic at one-dimensional quantum critical points.

Our results establish a direct connection between non-local magic and the spectral properties of the reduced density matrix, placing its study within the framework of entanglement spectroscopy and enabling analytical techniques developed for entanglement spectra to be directly applied~\cite{LiHaldane2008,Peschel2009,Calabrese2008}.
More importantly, the Walsh--Hadamard formulation identifies the autocorrelation structure of the entanglement spectrum as the fundamental quantity governing the Schmidt-gauge non-local magic. 
This perspective establishes a general analytical framework for investigating non-local magic in interacting many-body systems, quantum criticality, topological phases, and quantum dynamics.

%%%%%%%%%%%%%%%%%%%%%%%%%%%%%%%%%%%%%%%%%%%%%%%%%%%%%%%%%%%%%%%%%%%%%%%%%%%%%%
%%%%%%%%%%%%%%%%%%%%%%%%%%%%%%%%%%%%%%%%%%%%%%%%%%%%%%%%%%%%%%%%%%%%%%%%%%%%%%
%%%%%%%%%%%%%%%%%%%%%%%%%%%%%%%%%%%%%%%%%%%%%%%%%%%%%%%%%%%%%%%%%%%%%%%%%%%%%%
	
\textit{Schmidt-gauge non-local magic.—}	
We adopt the definition of non-local magic introduced in Ref.~\cite{Cao2025}, namely the minimum stabilizer Rényi entropy over all local unitary transformations, and specialize it to the second-order Rényi entropy.
Although this definition captures the genuinely non-local component of magic, its optimization rapidly becomes analytically intractable and numerically expensive with increasing system size.
Here we show that the Schmidt-gauge formulation admits an exact Walsh--Hadamard representation.
The analytical and numerical results presented below and in Ref.~\cite{associated} strongly support the conjecture that this formulation coincides with the non-local magic introduced in Ref.~\cite{Cao2025}.

Consider a pure state $|\psi\rangle$ of a system composed of $L$ qubits, partitioned into two subsystems $A$ and $B$, containing $m_A$ and $m_B$ qubits, respectively, with $m_A+m_B=L$.
Its Schmidt decomposition reads $\ket{\psi} = \sum_{\alpha=1}^{\chi} \sqrt{\lambda_\alpha} \ket{u_\alpha}_A \ket{v_\alpha}_B$, where $\{\ket{u_\alpha}_A\}$ and $\{\ket{v_\alpha}_B\}$ are orthonormal Schmidt bases, $\lambda_\alpha\ge0$ are the Schmidt eigenvalues, and
$\chi\le2^m$ is the Schmidt rank, with $m=\min(m_A,m_B)$.
Since the Schmidt vectors form orthonormal bases of the two subsystems, there always exist local unitary transformations that map them onto the computational basis. 
Ordering the Schmidt eigenvalues in descending order and padding the spectrum with zeros whenever necessary defines the canonical {\it Schmidt-gauge representative}:
\begin{equation}
	\ket{\psi}_{\rm Sch} \equiv \sum_{x\in\mathbb F_2^m} \sqrt{\lambda_x} \ket{x}_A \ket{x}_B,
	\label{psisch}
\end{equation}
where $\mathbb F_2^m$ denotes the $m$-dimensional vector space over the finite field $\mathbb F_2$, whose elements are binary strings \mbox{$x=(x_1,\ldots,x_m)$} with $x_i\in\{0,1\}$.

For a generic pure state of $L$ qubits, the second-order stabilizer Rényi entropy is defined as
\begin{equation}
	M_2(\ket{\phi}) \equiv -\log_2 \left[ 2^{-L} \sum_{P\in\mathcal P_L} \langle P\rangle_\phi^4 \right],
	\label{M2Def}
\end{equation}
where $\mathcal P_L$ denotes the set of phase-free Pauli strings. 
Applying Eq.~\eqref{M2Def} to the Schmidt-gauge representative in Eq.~\eqref{psisch}, one finds that only Pauli strings inducing the same binary action on the two Schmidt registers contribute to the Pauli sum.
In this way, the Schmidt-gauge state naturally defines
\begin{equation}
	M_2^{\rm Sch}(\psi) \!=\! M_2\!\left(\ket{\psi_{\rm Sch}} \right) \!=\! -\log_2 \left[ 2^{-m}\!\!\! \sum_{s,k\in\mathbb F_2^m} \!\!\!\! A_s(k)^4 \right]\!,
\label{eq:SchmidtGaugeMagic}
\end{equation}
where
\begin{equation}
	A_s(k) = \sum_{x\in\mathbb F_2^m} \sqrt{\lambda_x\lambda_{x\oplus s}} (-1)^{k\cdot x}.
\end{equation}
Here $x\oplus s$ denotes bitwise addition modulo two and $k\cdot x=\sum_{j=1}^m k_jx_j$ is the binary scalar product.
The details of the derivation can be found in~\cite{associated}. 
Since $M_2^{\rm NL}\leq M_2^{\rm Sch}$, Eq.~(\ref{eq:SchmidtGaugeMagic}) provides an upper bound to the optimized non-local magic. 
Extensive numerical evidence reported in Table~I strongly supports the saturation of this bound, while the equality has been proven analytically for $1\times L$ bipartitions~\cite{associated}.

Defining the autocorrelation function $f_s(x)=\sqrt{\lambda_x\lambda_{x\oplus s}}$ for each shift $s$, one immediately recognizes $A_s(k)$ as its Walsh--Hadamard transform~\cite{Beauchamp1975, Carlet2021}.
Equation~\eqref{eq:SchmidtGaugeMagic} therefore expresses non-local magic as the fourth Walsh moment of the binary autocorrelations of the ordered entanglement spectrum.
This representation is equivalent to the spectral expression derived in Eq.~(55) of Ref.~\cite{Cao2025}, however, the two formulations are not equally suited for analytical investigations. 
While Eq.~\eqref{eq:SchmidtGaugeMagic} immediately yields the spectral expression of Ref.~\cite{Cao2025},
the converse derivation requires uncovering hidden symmetry structures that are not manifest in the original representation. 
Making this hidden harmonic structure explicit is precisely what makes the analytical developments presented below possible.

\begin{table}[t]
	\caption{
	Summary of the numerical optimization testing the conjectured equivalence between the Schmidt-gauge non-local magic and the minimum second-order Stabilizer Rényi entropy over arbitrary local unitary transformations.
	For each system size and bipartition, the table reports the number of randomly generated states (Samples), the number of optimization runs ($N_{\mathrm{start}}$), and the number of optimization steps per run ($N_{\mathrm{steps}}$).
	Across all optimization runs, the minimum value obtained always satisfied
	$M_2 \geq M_2^{\rm Sch}(\psi)$, providing strong numerical evidence for the conjectured equivalence.}
	\label{tab:numerics}
	\begin{ruledtabular}
		\begin{tabular}{ccccc}
			$L$ & Bipartition & Samples & $N_{\mathrm{start}}$ & $N_{\mathrm{steps}}$  \\
			\hline
			2 & $1|1$ & 3600 & 500 & 5000 \\
			4 & $1|3$ & 3600 & 500 & 5000 \\
			4 & $2|2$ & 3600 & 500 & 5000 \\
			6 & $1|5$ & 1200 & 500 & 5000 \\
			6 & $2|4$ & 1200 & 500 & 5000 \\
			6 & $3|3$ & 1000 & 500 & 5000 
		\end{tabular}
	\end{ruledtabular}
\end{table}

%%%%%%%%%%%%%%%%%%%%%%%%%%%%%%%%%%%%%%%%%%%%%%%%%%%%%%%%%%%%%%%%%%%%%%%%%%%%%%
%%%%%%%%%%%%%%%%%%%%%%%%%%%%%%%%%%%%%%%%%%%%%%%%%%%%%%%%%%%%%%%%%%%%%%%%%%%%%%
%%%%%%%%%%%%%%%%%%%%%%%%%%%%%%%%%%%%%%%%%%%%%%%%%%%%%%%%%%%%%%%%%%%%%%%%%%%%%%

\textit{Spectral signatures of non-local magic.—}
Equation~\eqref{eq:SchmidtGaugeMagic} shows that the Schmidt-gauge non-local magic is entirely determined by the structure of the entanglement spectrum.
This formulation allows one to investigate analytically how different spectral structures determine the behavior of non-local magic.
We begin with one of the simplest possible cases, namely a perfectly flat distribution of the Schmidt eigenvalues, which corresponds to a maximally entangled state.
For such a state, all eigenvalues are identical, $\lambda_x=1/D$, with $D=2^m$.
Using the orthogonality of the Walsh characters, one immediately finds $A_s(k)=\delta_{k,0}$, independently of the shift $s$.
Consequently, $M_2^{\rm Sch}=0$.
This result is remarkable.
Although the state possesses the maximum possible bipartite entanglement, its Schmidt-gauge non-local magic vanishes identically, demonstrating that non-local magic is governed by the structure of the entanglement spectrum rather than by the amount of entanglement alone.

Equation~(\ref{eq:SchmidtGaugeMagic}) provides a perturbative estimate for Haar-random states~\cite{Page1993}. 
We write $\lambda_x=(1+\eta_x)/D$, with $\sum_x\eta_x=0$ and $\sigma^2=D^{-1}\sum_x\eta_x^2$. 
Within this approximation, the fluctuations $\eta_x$ are treated as weak and nearly uncorrelated; for balanced bipartitions, this assumption should be regarded as heuristic after monotonic Schmidt ordering. 
Accordingly, the autocorrelations $f_s(x)=\sqrt{\lambda_x\lambda_{x\oplus s}}$ remain nearly uniform and their Walsh--Hadamard transforms are dominated by the $k=0$ component. Expanding to second order gives $A_s(0)\simeq1-(\sigma^2-C_s)/4$, where $C_s=D^{-1}\sum_x\eta_x\eta_{x\oplus s}$.
For generic nonzero shifts, decorrelation implies $C_s\simeq0$, whereas $A_0(0)=1$ exactly, producing only an $O(D^{-1})$ correction to the normalized fourth moment. 
The nonzero Walsh modes are suppressed by random cancellations, $A_s(k\neq0)=O(\sigma/\sqrt D)$, yielding $M_2^{\rm Sch}(\psi)\simeq-4\log_2(1-\sigma^2/4)+O(D^{-1})$. 
Consequently, within this approximation, the Schmidt-gauge non-local magic remains finite in the thermodynamic limit despite the volume-law scaling of the entanglement entropy.

%%%%%%%%%%%%%%%%%%%%%%%%%%%%%%%%%%%%%%%%%%%%%%%%%%%%%%%%%%%%%%%%%%%%%%%%%%%%%%
%%%%%%%%%%%%%%%%%%%%%%%%%%%%%%%%%%%%%%%%%%%%%%%%%%%%%%%%%%%%%%%%%%%%%%%%%%%%%%
%%%%%%%%%%%%%%%%%%%%%%%%%%%%%%%%%%%%%%%%%%%%%%%%%%%%%%%%%%%%%%%%%%%%%%%%%%%%%%

\begin{figure}
	\begin{center}
		\includegraphics[width=0.9\columnwidth]{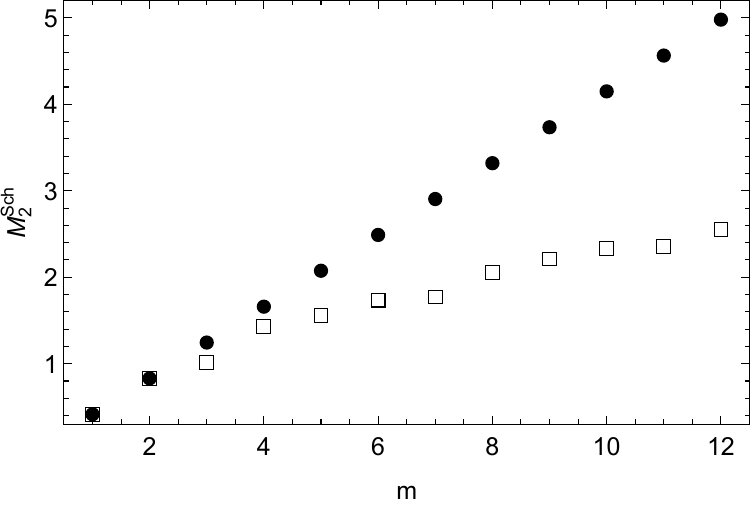}
	\end{center}
	\caption{Comparison between the non-local magic associated with the natural factorized ordering of the entanglement spectrum~\cite{Collura2026,Iannotti} (filled circles) and that obtained after monotonically ordering the Schmidt coefficients (open squares). We consider spectra generated by $m$ weakly non-degenerate binary modes, with $\theta_j=(\pi/8)(1+0.02\, r_j)$ and $r_j$ uniformly sampled in ([-1,1]). 
	The natural tensor-product ordering yields the additive result $M_2^{\rm fact}=\sum_{j=1}^{m}g(\nu_j)$, whereas the monotonically ordered Schmidt representative systematically gives smaller values.}
	\label{Figura1}
\end{figure}

\textit{Factorized entanglement spectra.—}
The previous examples show that non-local magic is governed by the structure of the entanglement spectrum rather than by the amount of entanglement alone.
This naturally motivates the study of spectra with additional algebraic structure.
Recently, Refs.~\cite{Collura2026,Iannotti} showed that, for Gaussian fermionic states, the canonical Gaussian representative yields an additive expression for non-local magic, leading to an extensive scaling for random Gaussian states and a logarithmic growth at one-dimensional criticality.
This result follows directly from the Walsh--Hadamard representation derived above.
Consider an entanglement spectrum whose eigenvalues admit the factorized representation $\tilde{\lambda}_{\mathbf{x}}=\prod_{j=1}^{m}(1+(-1)^{x_j}\nu_j)/2$, where $\mathbf{x}=(x_1,\ldots,x_m)$ labels the binary configurations and the parameters $\nu_j$ are sorted in decreasing order.
With this tensor-product labelling, Eq.~\eqref{eq:SchmidtGaugeMagic} reduces to $M_2^{\rm fact}=\sum_{j=1}^{m}g(\nu_j)$, where
\begin{equation}
	g(\nu)=-\log_2\!\left(1-\nu^2+\nu^4\right).
	\label{spdef}
\end{equation}
Up to the convention for the logarithm base, this coincides with the expression of Refs.~\cite{Collura2026,Iannotti}. 
The corresponding second-order Rényi entropy is likewise additive, $S_2^{\rm fact}=\sum_{j=1}^{m}s(\nu_j)$, where $s(\nu)=-\log_2[(1+\nu^2)/2]$.

However, for $m\ge3$ the resulting ordering generally differs from the monotonic Schmidt ordering adopted in the present work. 
Consequently, the Schmidt-gauge non-local magic no longer admits an additive decomposition, since Walsh autocorrelations are not invariant under arbitrary permutations of the Schmidt eigenvalues.
Numerical calculations consistently show that the Schmidt representative yields values smaller than $M_2^{\rm fact}$ (Fig.~\ref{Figura1}), with equality only for $m\le2$, where the two ordering prescriptions coincide.
A possible interpretation is that the canonical Gaussian representative naturally belongs to the local Gaussian orbit, whereas the Schmidt representative may require local basis transformations outside this manifold.
If so, $M_2^{\rm fact}$ characterizes the optimum over local Gaussian transformations, while the Schmidt-gauge non-local magic corresponds to the unrestricted local-unitary optimization.
Establishing whether this interpretation is exact remains an interesting open problem.

Since the Schmidt representation no longer admits an additive decomposition analogous to $M_2^{\rm fact}$, it is no longer obvious whether the Schmidt-gauge non-local magic can become extensive in the thermodynamic limit.
The systematic reduction observed numerically indicates that monotonic Schmidt ordering plays a non-trivial role in determining the asymptotic scaling.

%%%%%%%%%%%%%%%%%%%%%%%%%%%%%%%%%%%%%%%%%%%%%%%%%%%%%%%%%%%%%%%%%%%%%%%%%%%%%%
%%%%%%%%%%%%%%%%%%%%%%%%%%%%%%%%%%%%%%%%%%%%%%%%%%%%%%%%%%%%%%%%%%%%%%%%%%%%%%
%%%%%%%%%%%%%%%%%%%%%%%%%%%%%%%%%%%%%%%%%%%%%%%%%%%%%%%%%%%%%%%%%%%%%%%%%%%%%%

\textit{Non-local magic versus entanglement.—}
The Walsh--Hadamard formulation also makes it possible to establish a direct quantitative relation between non-local magic and entanglement. 
In particular, one can prove the following universal bound:
\begin{equation}
	M_2^{\rm Sch}(\psi)\le 2S_2(\rho_A),
	\label{eq:bound}
\end{equation}
where $S_2(\rho_A) = -\log_2\!\left(\sum_x\lambda_x^2\right)$ denotes the second-order Rényi entanglement entropy~\cite{Eisert2010}.
The proof is remarkably simple.
Starting from the observation $\sum_{s,k}A_s(k)^4\ge\sum_kA_0(k)^4$, where
$A_0(k)=\sum_x\lambda_x(-1)^{k\cdot x}$ is the Walsh--Hadamard transform of the entanglement spectrum, the Cauchy--Schwarz inequality gives $\sum_kA_0(k)^4\ge2^{-m}\! \left(\sum_kA_0(k)^2\right)^2$, while Parseval's identity yields $\sum_kA_0(k)^2= 2^m\sum_x\lambda_x^2$.
Substituting these relations into Eq.~\eqref{eq:SchmidtGaugeMagic} immediately gives Eq.~\eqref{eq:bound}.

Equation~\eqref{eq:bound} identifies extensive entanglement as a necessary, but not sufficient, condition for extensive non-local magic, for extensive non-local magic, as demonstrated by maximally entangled states, for which $S_2=m$ while \mbox{$M_2^{\rm Sch}=0$}.
This bound also reveals that Schmidt-gauge non-local magic probes finer features of the entanglement spectrum than the Rényi entropy alone, since the latter depends only on the distribution of the Schmidt eigenvalues, whereas the former is additionally sensitive to the binary correlation structure encoded in their Walsh autocorrelations.

%%%%%%%%%%%%%%%%%%%%%%%%%%%%%%%%%%%%%%%%%%%%%%%%%%%%%%%%%%%%%%%%%%%%%%%%%%%%%%
%%%%%%%%%%%%%%%%%%%%%%%%%%%%%%%%%%%%%%%%%%%%%%%%%%%%%%%%%%%%%%%%%%%%%%%%%%%%%%
%%%%%%%%%%%%%%%%%%%%%%%%%%%%%%%%%%%%%%%%%%%%%%%%%%%%%%%%%%%%%%%%%%%%%%%%%%%%%%

\textit{Ground states of local Hamiltonians.—}
The bound in Eq.~\eqref{eq:bound} immediately constrains the scaling of Schmidt-gauge non-local magic in physically relevant many-body systems.
More generally, whenever the second-order Rényi entropy satisfies an area law, Schmidt-gauge non-local magic is bounded by the same scaling.
In particular, for one-dimensional gapped Hamiltonians, where $S_2(\rho_A)=O(1)$~\cite{Vidal2003,Amico2008}, Eq.~\eqref{eq:bound} implies $M_2^{\rm Sch}(\psi)=O(1)$, showing that Schmidt-gauge non-local magic remains finite throughout a gapped phase.

The situation is qualitatively different for one-dimensional quantum critical systems, where the second-order Rényi entropy grows logarithmically with the subsystem size.
An important benchmark is provided by one-dimensional free-fermion Hamiltonians, whose reduced density matrix factorizes into independent entanglement modes~\cite{Peschel2003},
$\rho_A=\bigotimes_{j=1}^{m}\rho_j$, with occupation probabilities $p_j=(1+e^{\varepsilon_j})^{-1}$, where $\varepsilon_j$ are the single-particle entanglement energies.
Chung and Peschel~\cite{Peschel2001} showed that, at criticality, the entanglement Hamiltonian consists of approximately equally spaced single-particle levels, with spacing $\varepsilon_L\propto1/ \log L$.
For the corresponding tensor-product (Gaussian) representative, one finds $M_2^{\rm G}= \kappa\log_2L+O(1)$, where $\kappa=\frac1\pi\int_0^\infty g[(1+e^z)^{-1}]\,dz$  with $g(\nu)$ defined in Eq.~\eqref{spdef}. 

\begin{figure}
	\begin{center}
		\includegraphics[width=0.9\columnwidth]{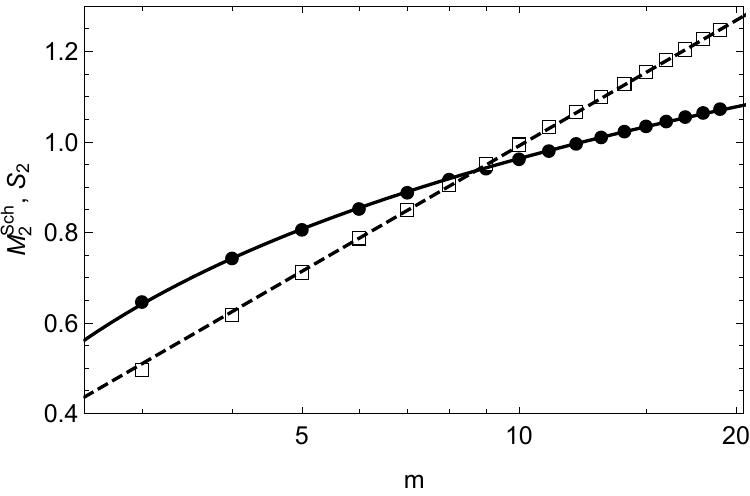}
	\end{center}
\caption{
	Schmidt-gauge non-local magic $M_2^{\rm Sch}$ (black dots) and second-order Rényi entanglement entropy $S_2$ (white squares) computed from the ordered Calabrese--Lefevre spectrum as a function of the subsystem size $m$.
	For each $m$, the reduced density-matrix spectrum is constructed by sampling the analytical Calabrese--Lefevre cumulative distribution at equally spaced probabilities $(k-\tfrac12)/2^m$ ($k=1,\ldots,2^m$) and subsequently sorting the resulting eigenvalues in decreasing order.
	The solid and dashed curves correspond to the best fits of $M_2^{\rm Sch}$ and $S_2$, respectively, $M_2^{\rm Sch}(m)\simeq A+B\ln\ln m$ and $S_2(m)\simeq C+0.4\ln m$.
	Note that this behavior strongly depends on the scaling of the parameter $b$ in the Calabrese-Lefevre formula~\cite{Calabrese2008}. 
	For instance, if $b$ were taken independent from $m$, $M_2^{\rm Sch}$ would saturate to a finite value.
}
\label{Figura2}
\end{figure}

The Stabilizer Rényi entropy of the tensor-product Gaussian representative, $M_2^{\rm G}$, does not, in general, coincide with the Schmidt-gauge value $M_2^{\rm Sch}$.
Sorting the many-body eigenvalues in decreasing order removes the factorization into independent modes and can substantially change their Walsh--Hadamard correlations. 
To investigate this effect, we construct the ordered spectrum from the universal Calabrese--Lefevre distribution~\cite{Calabrese2008}. 
Introducing $ b=-\ln\lambda_{\max}$, the cumulative number of eigenvalues larger than $\lambda$ is equal to $ n(\lambda)= I_0\!\left( 2\sqrt{ b\ln(\lambda_{\max}/\lambda)} \right)$, where $I_0$ denotes the modified Bessel function of the first kind of order zero.
At a conformal critical point, where $b=(c/6)\ln L+O(1)$, the large-argument expansion of the modified Bessel function yields the asymptotic Schmidt spectrum \mbox{$\lambda_r \sim \exp\!\left[-b-(\ln r)^2/(4b)\right]$}.
This suggests a natural organization of the ordered spectrum into dyadic shells, i.e., rank intervals bounded by successive powers of two $(2^q \le r \le 2^{q+1})$, which are naturally adapted to the binary translations entering the Walsh--Hadamard representation. 
The total spectral weight contained in the $q$-th shell behaves as 
$\mu_q \simeq \exp\!\left[-(q\ln2-2b)^2/(4b)\right]$, so that the relevant Walsh structure is distributed over a number of binary scales proportional to  $\Delta q\sim\sqrt{b}\sim\sqrt{\ln L}$.
If the normalized fourth Walsh moment scales algebraically as $b^{-\alpha}$, we recover
$M_2^{\rm Sch} =\alpha\log_2\!\ln L+O(1)$.
The numerical results shown in Fig.~\ref{Figura2} support this scaling scenario.
For the Calabrese--Lefevre spectra accessible to our calculation, \mbox{$m\le20$}, $M_2^{\rm Sch}$ continues to increase, but much more slowly than $M_2^{\rm G}$.
In particular, over the accessible range,
\begin{equation}
	M_2^{\rm Sch}(m) \simeq
	A+B\ln\ln m,
	\label{eq:numericalDoubleLog}
\end{equation}
with $A\simeq0.60$ and $B\simeq0.43$, yielding a root-mean-square error of approximately $2\times10^{-3}$.
Although these system sizes are not sufficient to establish the asymptotic coefficient or to exclude an eventual saturation with exceptionally slow corrections, they clearly show
that the ordered Schmidt representative does not follow the $\log L$ growth of the tensor-product Gaussian representative.
The suppression of Schmidt-gauge non-local magic therefore originates entirely from the reordering of the same entanglement spectrum induced by Schmidt ordering and the resulting modification of its Walsh binary correlations.

%%%%%%%%%%%%%%%%%%%%%%%%%%%%%%%%%%%%%%%%%%%%%%%%%%%%%%%%%%%%%%%%%%%%%%%%%%%%%%
%%%%%%%%%%%%%%%%%%%%%%%%%%%%%%%%%%%%%%%%%%%%%%%%%%%%%%%%%%%%%%%%%%%%%%%%%%%%%%
%%%%%%%%%%%%%%%%%%%%%%%%%%%%%%%%%%%%%%%%%%%%%%%%%%%%%%%%%%%%%%%%%%%%%%%%%%%%%%

\textit{Conclusions.—}
Building on the Schmidt-gauge formulation introduced in Ref.~\cite{Cao2025}, we have shown that the corresponding non-local magic admits an equivalent representation as the fourth Walsh moment of the Walsh--Hadamard autocorrelations of the entanglement spectrum.
This representation makes the underlying harmonic structure explicit and provides a natural analytical framework for investigating non-local magic.

Within this framework, we derived exact analytical results for broad classes of quantum states, revealing how different structures of the entanglement spectrum generate qualitatively different behaviors of non-local magic.
We also proved the universal bound $M_2^{\rm Sch}\le2S_2$, showing that the scaling of Schmidt-gauge non-local magic is fundamentally constrained by the scaling of the second-order Rényi entanglement entropy.
As an immediate consequence, Schmidt-gauge non-local magic satisfies an area law whenever the second-order Rényi entanglement entropy does.

The Walsh--Hadamard formulation also clarifies the role of factorized entanglement spectra.
It naturally reproduces the additive expression recently obtained for the canonical Gaussian representative~\cite{Collura2026,Iannotti}, while showing that the monotonic Schmidt ordering generally breaks this factorization.
An immediate consequence is that, while entanglement preserves the quasi-particle factorization of Gaussian (or integrable) systems, Schmidt-gauge non-local magic probes additional correlations between the entanglement modes that remain invisible to conventional entanglement measures.

\begin{table}[t]
	\centering
	\begin{tabular}{lcc}
		\hline\hline
		State family &
		$S_2$ &
		$M_2^{\rm Sch}$ \\
		\hline
		Maximally entangled
		& $m$ 
		& $0$ \\
		
		Haar-random
		& $\simeq m$
		& $O(1)$ \\
			
		1D gapped ground states
		& $O(1)$
		& $O(1)$ \\
		
		1D critical systems
		& $\frac{c}{4}\log m$ \quad \qquad
		& $ \simeq \kappa \log\!\log m$ \\
		\hline\hline
	\end{tabular}
	\caption{Comparison between the scaling of the second-order Rényi entanglement entropy and the Schmidt-gauge non-local magic for the different classes of states discussed in the text. In all cases, the scaling is reported as a function of the number $m$ of qubits in the smaller subsystem.}	\label{tab:summary}
\end{table}

Finally, our analytical arguments suggest that the Schmidt ordering suppresses the logarithmic growth expected from the entanglement spectrum, leading instead to a much slower, possibly double-logarithmic, scaling.
The numerical results are in excellent agreement with this prediction over the accessible range of subsystem sizes.
Whether this mechanism universally governs the asymptotic scaling of Schmidt-gauge non-local magic in critical systems, and how it extends to volume-law states, remains an important open question.

More generally, the Walsh--Hadamard formulation changes the analytical perspective on non-local magic.
A quantity originally defined through a highly non-trivial optimization becomes directly accessible through the harmonic structure of the entanglement spectrum.
This establishes entanglement spectroscopy as a natural analytical framework for investigating non-local magic and opens the way to analytical studies of interacting systems, higher-dimensional phases, topological matter, nonequilibrium dynamics, and mixed states.

\textit{Acknowledgements.—}
G.T. is supported by the Slovenian Quantum Science Hub — SQUASH programme, co-funded by the European Union under the Horizon Europe Marie Skłodowska-Curie Actions COFUND programme, grant agreement No. 101177446, and by the Slovenian Research and Innovation Agency (ARIS), contract No. 5110-18/2025-5. SMG and FF acknowledge support from the project "Implementation of cutting-edge research and its application as part of the Scientific Center of Excellence for Quantum and Complex Systems, and Representations of Lie Algebras", Grant No. PK.1.1.10.0004, co-financed by the European Union through the European Regional Development Fund - Competitiveness and Cohesion Programme 2021-2027 and from the Croatian Science Foundation (HrZZ) through the project IP-2025-02-1667, Mining the Quantum: Frustration, Disorder, and Devices.

\clearpage
\onecolumngrid

\appendix

\section*{Supplementary Material -- Equivalence with the spectral representation of Ref.~\cite{Cao2025}}
\label{sec:equivalence_cao}

\setcounter{equation}{0}
\renewcommand{\theequation}{S\arabic{equation}}

We now prove that the Walsh--Hadamard representation derived in the main text is exactly equivalent to the spectral expression reported in Eq.~(55) of Ref.~\cite{Cao2025}. 
Throughout this section we consider a pure state of a bipartite system $A\cup B$, where the two subsystems contain $m_A$ and $m_B$ qubits, respectively, and we define $m=\min(m_A,m_B)$.
Using our notation, Eq.~(55) of Ref.~\cite{Cao2025} reads
\begin{equation}
	M_2^{\mathrm{Sch}}(\psi)= -\log_2\Bigg[ \sum_{i_1,i_2,i_3,i_4\in\mathbb F_2^m} \sqrt{ \lambda_{i_1} \lambda_{i_2} \lambda_{i_3} \lambda_{i_4} \lambda_{i_1\oplus i_2\oplus i_3} \lambda_{i_1\oplus i_2\oplus i_4} \lambda_{i_1\oplus i_3\oplus i_4} \lambda_{i_2\oplus i_3\oplus i_4} } \Bigg].
	\label{eq:cao_spectral_formula}
\end{equation}
On the other hand, our Walsh--Hadamard representation reads
\begin{equation}
	M_2^{\mathrm{Sch}}(\psi) = -\log_2\left[\mathcal W\right],
	\label{eq:SM_walsh_formula}
\end{equation}
where
\begin{equation}
	\mathcal W = 2^{-m} \sum_{s,k\in\mathbb F_2^m} A_s(k)^4, \qquad
	A_s(k) = \sum_{x\in\mathbb F_2^m} (-1)^{k\cdot x} \sqrt{\lambda_x\lambda_{x\oplus s}}
	\label{eq:walsh_functional_equivalence}
\end{equation}
is the Walsh--Hadamard transform of the overlap function introduced in the main text.
Expanding the fourth power and interchanging the order of summation gives
\begin{equation}
	\mathcal W = 2^{-m} \sum_{s,k} \sum_{x_1,x_2,x_3,x_4} = (-1)^{k\cdot(x_1\oplus x_2\oplus x_3\oplus x_4)}
	\prod_{a=1}^{4} \sqrt{\lambda_{x_a}\lambda_{x_a\oplus s}}.
	\label{eq:fourth_walsh_expansion}
\end{equation}
The sum over the Walsh index can now be performed by using the orthogonality relation
\begin{equation}
	2^{-m} \sum_{k\in\mathbb F_2^m} (-1)^{k\cdot u} = \delta_{u,0},
	\label{eq:walsh_character_orthogonality}
\end{equation}
which yields
\begin{equation}
	\mathcal W = \sum_s \sum_{x_1,x_2,x_3,x_4} \delta_{x_1\oplus x_2\oplus x_3\oplus x_4,0} \prod_{a=1}^{4} \sqrt{\lambda_{x_a}\lambda_{x_a\oplus s}}.
	\label{eq:walsh_xor_constraint}
\end{equation}
The delta function imposes the XOR constraint $x_4=x_1\oplus x_2\oplus x_3$, reducing the quadruple sum to
\begin{equation}
	\mathcal W = \sum_{s,x_1,x_2,x_3} \sqrt{ \lambda_{x_1} \lambda_{x_2} \lambda_{x_3} 	\lambda_{x_1\oplus x_2\oplus x_3} 	\lambda_{x_1\oplus s} \lambda_{x_2\oplus s} \lambda_{x_3\oplus s} \lambda_{x_1\oplus x_2\oplus x_3\oplus s} }.
	\label{eq:walsh_reduced_sum}
\end{equation} 
We now introduce the linear change of variables $y_i=x_i$ for $i=1,\cdot,3$ and $y_4=x_1\oplus x_2\oplus x_3\oplus s$
Since addition in $\mathbb F_2^m$ is invertible, this transformation is bijective.
The shifted indices then become 
\begin{align}
	x_1\oplus s & = y_2\oplus y_3\oplus y_4, \nonumber \\
	x_2\oplus s & = y_1\oplus y_3\oplus y_4, \nonumber \\
	x_3\oplus s & = y_1\oplus y_2\oplus y_4,  \\
	x_1\oplus x_2\oplus x_3\oplus s	& = y_4. \nonumber
\end{align}
Substituting these identities into Eq.~\eqref{eq:walsh_reduced_sum} gives
\begin{equation}
	\mathcal W = \sum_{y_1,y_2,y_3,y_4} \sqrt{ \lambda_{y_1} \lambda_{y_2} \lambda_{y_3} \lambda_{y_1\oplus y_2\oplus y_3} \lambda_{y_2\oplus y_3\oplus y_4} \lambda_{y_1\oplus y_3\oplus y_4} 	\lambda_{y_1\oplus y_2\oplus y_4} \lambda_{i_4} }.
	\label{eq:walsh_cao_final}
\end{equation}
This expression coincides exactly with the argument of the logarithm in Eq.~\eqref{eq:cao_spectral_formula}. Therefore,
\begin{equation}
	2^{-m} \sum_{s,k\in\mathbb F_2^m} A_s(k)^4 = \!\!\!\! \sum_{y_1,y_2,y_3,y_4\in\mathbb F_2^m}
	\sqrt{ \lambda_{y_1} \lambda_{y_2} \lambda_{y_3} \lambda_{y_4} 	\lambda_{y_1\oplus y_2\oplus y_3} \lambda_{y_1\oplus y_2\oplus y_4} \lambda_{y_1\oplus y_3\oplus y_4} \lambda_{y_2\oplus y_3\oplus y_4}},
	\label{eq:exact_equivalence}
\end{equation}
which proves that the Walsh--Hadamard representation derived in the main text and the spectral representation of Ref.~\cite{Cao2025} are mathematically identical.
	
\end{document}